\documentstyle[12pt]{article}
\begin{document}
\begin{titlepage}
\title{Symmetries of Nonrelativistic Phase Space\\
and the Structure of Quark-Lepton Generation
\footnote{Talk given
during the DICE2008 conference, September 2008, Castiglioncello,
Italy}}
\author{
{Piotr \.Zenczykowski }\\
{\em Division of Theoretical Physics},
{\em Institute of Nuclear Physics,}\\
{\em Polish Academy of Sciences,}\\
{\em Radzikowskiego 152,
31-342 Krak\'ow, Poland}\\
E-mail: piotr.zenczykowski@.ifj.edu.pl
}
\maketitle
\begin{abstract}
According to the Hamiltonian formalism, nonrelativistic
phase space may be considered as
an arena of physics, with momentum and position treated as 
independent variables.
Invariance of ${\bf x}^2+{\bf p}^2$
constitutes then a natural generalization of ordinary rotational invariance.
We consider
Dirac-like linearization of this form, with position and momentum satisfying
standard commutation relations. This leads to
the identification of a quantum-level structure 
from which some phase space properties might emerge. 

Genuine rotations and reflections in phase space are tied
to the existence of new quantum numbers, unrelated to ordinary 3D space.
Their properties allow their
identification with
 the internal quantum
numbers characterising the structure of
a single quark-lepton generation in the Standard Model. 
In particular,
 the algebraic
structure of the Harari-Shupe preon model of fundamental particles
is reproduced exactly and without invoking any sub-particles.

Analysis of the Clifford algebra of nonrelativistic phase space
singles out an element 
which might be associated with the 
concept of lepton mass. This element is transformed into a
corresponding element for a single coloured quark, leading to a generalization
of the concept of mass and a different starting point
for the discussion of quark unobservability.
\end{abstract}
\vfill
\end{titlepage}

\begin{flushright}
{\parbox{300pt}{\it ``I do not believe that a real understanding of the nature of
elementary particles can ever be achieved without a simultaneous deeper
understanding of the nature of spacetime''
\begin{flushright}Roger Penrose - \cite{Penrose}\end{flushright}}}
\end{flushright}

\section{Elementary particles and space}
In our search for the underlying components of matter we have identified several
species of fundamental elementary fermions. According to the Standard Model, there are three
generations of such particles, each generation composed of eight objects: 
two leptons and two
triplets of quarks. Interactions of these particles proceed through an
exchange of various gauge bosons. 
The particles themselves (and systems composed thereof)
 differ in their
properties, with the differences corresponding to different eigenvalues of various
quantum numbers.

Two groups of such quantum numbers may be identified. The first one is composed of
the so-called ``spatial'' quantum numbers, the other one comprises ``internal'' quantum
numbers.

The spatial quantum numbers of a particle are those standardly written in the form
``$J^{PC}$''. Here $J$ denotes the spin of the particle, $P$ denotes its parity, 
and $C$ is the so-called charge-conjugation parity. 
They are all connected with the
properties of spacetime: spin is related  to ordinary rotations, parity - to 
reflections in our 3D space, while $C$-parity - being related to 
complex conjugation - is clearly connected with time reflection.
It is important to notice that all these spatial quantum numbers are
nonrelativistic in origin. In particular, the emergence of
antiparticles is not a relativistic phenomenon as the Dirac equation 
might suggest: they appear also when the
strictly nonrelativistic Schr\"odinger equation is linearized \cite{HK2003}.

As far as internal quantum numbers are concerned,
 many physicists 
believe
  that these quantum numbers should also be connected with
 some ``classical arena'' which would constitute some kind of an extension of
 spacetime.
 In view of the nonrelativistic nature of all spatial quantum numbers
it seems then natural to expect that the minimal
extension of the concept of space needed for an understanding of internal
quantum numbers should be nonrelativistic as well.

\section{Space, time and quantum}
It might be argued that the reasoning of the preceding section, based on 
an expected analogy between spatial and internal
quantum numbers, is not sufficient to justify
 the adoption of
a nonrelativistic
approach well enough.
After all, the standard form of the theory of special relativity involves 
transformations  which mix time with space, thereby undermining the very concept of absolute
simultaneity.

In fact, however,
 the connection between space and time is more subtle than 
   the standard form of special relativity would
suggest: 
the latter form emerges only when the Einstein radiolocation prescription 
for the synchronization of distant clocks is adopted. 
Yet, distant clocks may be
synchronised in various ways, reflecting the presence of gauge
freedom related to the impossibility of measuring the one-way speed of light.
 Absolute simultaneity may then be achieved with the help of a suitable gauge
\cite{Mansouri}.

A different argument in favour of a nonrelativistic approach may also be
given. Namely, 
although relativistic field theory does unite special relativity and quantum
physics,  this marriage of quantum
and relativistic ideas is in the opinion of many physicists 
somewhat uneasy \cite{Bell}. The actual wording takes various forms, such as
e.g. 
``The construction of a fully objective theory of state-vector reduction which
is consistent with the spirit of relativity is a profound challenge, since
`simultaneity' is a concept (...) foreign to relativity''\cite{Emperor}.
 Considerations of this type lead to a widely advocated idea that space and time are
  emergent phenomena,
absent at the underlying quantum level.

We conclude therefore that it is well justified to adopt an 
approach in which one does not {\it start} from mixing time with space.
 Yet, if the
internal quantum numbers are to be connected with the 
properties of the macroscopic
``space", the ordinary 3D
space clearly has to be somehow extended into a broader ``arena". 

\section{The arena}
\subsection{Max Born: a hint}
The issue of a possible relation between some particle properties, such as mass, 
and the surrounding ``emergent'' space was of significant concern 
already to Max Born,
over half a century ago. In his 1949 paper \cite{Born1949}
he writes: ``I think that the assumption of the observability of the
4-dimensional distance of two events inside atomic dimensions (no clocks or
measuring rods) is an extrapolation..."
He then continues with the discussion of a difference between the position and
momentum spaces for elementary particles. First, he
notes that the concept of mass appears in the relation
$p^2=m^2$, and that different observed elementary particles correspond
 to different discrete values
 of $m^2$, thus rendering $p^2$ observable.  
 Then, he stresses that
$x^2$, the corresponding invariant in coordinate space 
(with $x^2$ of atomic dimensions),
seems to be ``no observable at all".

At the same time, he points out that the laws of nature such as
\begin{eqnarray}
&&\dot{x}_k=\frac{\partial H}{\partial p_k},~~~
\dot{p}_k=-\frac{\partial H}{\partial x_k},\nonumber\\
&&{}[x_k,p_l]=i\hbar\delta_{kl},\nonumber\\
&&{}L_{kl}=x_kp_l-x_lp_k
\end{eqnarray}
are invariant under ``reciprocity" transformations:
\begin{eqnarray}
x_k \to p_k,&&p_k \to -x_k.
\end{eqnarray}
 Noting that the reciprocity symmetry somehow does not apply to elementary
 particles, he concludes: ``This lack of symmetry seems to me very strange and
 rather improbable".

\subsection{Phase space as the arena}
In view of the preceding arguments and with quantum mechanics ``living in phase
space", it seems natural to start from mixing the 3D space of positions with the
3D space of momenta, and to look for any additional quantum numbers that might
possibly emerge in such a broader scheme.
In other words, instead of implicitly identifying 
the ordinary 3D space of positions with
the  nonrelativistic arena
on which physical processes take
place,  I propose that it is
the nonrelativistic phase space that might 
and should be viewed as such an arena.

It should be stressed that the procedure of mixing the spaces
of positions and momenta leads to such a generalization of the concept of
ordinary space, which could be termed ``minimal''. 
In particular no additional dimensions - ``hidden'' from our sight -
are introduced in this way. 
We just have to recognize that this macroscopic arena is around us, fully
visible.
 
In the following, I will infer the existence of some internal quantum numbers
from the properties of this generalized arena. 
In a more fundamental approach, however, 
one has to reverse the arrow of implications and -
starting
from the underlying quantum structure, as manifested in quantum
numbers of elementary particles -
actually build the ``Emergent
Phase Space" (EPS). 

\subsection{Phase space and the Standard Model}
The simplest phase-space
generalization of the 3D concepts of reflection and rotation
requires a fully symmetric treatment of the two O(3) invariants,
${\bf x}^2$ and ${\bf p}^2$, which is achieved by adding them together:
\begin{eqnarray}
{\bf x}^2+{\bf p}^2
\end{eqnarray}
This generalization leads to $O(6)$, which goes beyond both Born's reciprocity and the familiar 
symmetries of 3D space. 
Treating ${\bf x}^2$ and ${\bf p}^2$ as operators we now
require their commutators
to be form invariant. As is well known from the case of 
the 3D harmonic oscillator the original $O(6)$ symmetry
is then reduced to $U(1)\otimes SU(3)$. 
The $U(1)$ factor describes (in particular) 
Born's reciprocity transformations and their
squares: 3D reflections.
The $SU(3)$ factor takes care of standard rotations
(among other transformations).

The appearance of the $U(1)\otimes SU(3)$ group from the first principles
and the presence of the same group in the Standard Model (SM)
raises the possibility that 
the SM internal symmetry group 
is actually related to phase space symmetries.
A confirmation of this suggestion seems to require
the construction of the SM gauge prescription 
from and/or upon the underlying quantum structure. 
The gauge structure would have to appear alongside
 the emerging (phase)
space. Consequently, I think it lies beyond our reach at the moment.
In the following, I shall show, however, that 
the structure of quantum numbers obtained at the quantum level of
 the phase-space-related approach exactly parallels 
 that observed in the real world. 

\section{Linearization of ${\bf x}^2+{\bf p}^2$}
In order to reach the quantum level, we linearize ${\bf x}^2+{\bf p}^2$
\`a la Dirac. Using anticommuting $A_k$ and $B_k$ ($k=1,2,3$) , with explicite
representation
\begin{eqnarray}
A_k&=&\sigma_k\otimes\sigma_0\otimes\sigma_1\nonumber\\
B_k&=&\sigma_0\otimes\sigma_k\otimes\sigma_2\\
B_7&=&\sigma_0\otimes\sigma_0\otimes\sigma_3\nonumber
\end{eqnarray}
($B_7$ is the seventh anticommuting element of the relevant 
Clifford algebra),
one finds
\begin{equation}
({\bf A}\cdot {\bf p}+{\bf B}\cdot {\bf x})
({\bf A}\cdot {\bf p}+{\bf B}\cdot {\bf x})=
({\bf p}^2+{\bf x}^2)+ \sum_1^3 \sigma_k\otimes\sigma_k\otimes\sigma_3.
\end{equation}
The first term on the r.h.s., denoted below by $R$, appears here because 
all six elements $A_k$ and $B_l$ anticommute among themselves.
The second term, denoted by $R^{\sigma}$, 
is due to the fact that $x_k$ and $p_k$ do not commute.
These two terms sum up to a total $R^{tot}=R+R^{\sigma}$.

The $SU(4)/SO(6)$ generators are constructed as antisymmetric bilinears of
$A_k,B_l$.
In particular, the generator of standard rotation has the explicit form
\begin{equation}
S_k=\frac{1}{2}(\sigma_k\otimes\sigma_0+\sigma_0\otimes\sigma_k)\otimes\sigma_0
\end{equation}
and corresponds to simultaneous (same size and sense) rotations in momentum and
position subspaces.

\section{Eigenvalues of $R^{tot}$}
\subsection{Gell-Mann-Nishijima relation}
We now introduce operator $Y$, to be identified shortly with the (weak)
hypercharge:
\begin{equation}
Y\equiv \frac{1}{3}R^{\sigma}B_7=\frac{1}{3}\sum_1^3
\sigma_k\otimes\sigma_k\otimes\sigma_0\equiv 
\sum_1^3Y_k.
\end{equation}
Since the ``partial hypercharges'' $Y_k$ commute among themselves, they may be
simultaneously diagonalized. One then gets the pattern shown in Table
\ref{table2} (as the matrices are $8 \times 8$, this pattern is obtained twice).

\begin{table}
\caption{Decomposition of eigenvalue of Y into eigenvalues of its
components}
\label{table2}
\begin{center}
\begin{math}
\begin{array}{c|cccc}
{\rm colour~~}  & ~0 & ~1 & ~2 & ~3 \\
\hline 
Y &-1&+\frac{1}{3}&+\frac{1}{3}&+\frac{1}{3}
\rule{0mm}{6mm}\\
Y_1 &-\frac{1}{3}&-\frac{1}{3}&+\frac{1}{3}&+\frac{1}{3}
\rule{0mm}{6mm}\\
Y_2 &-\frac{1}{3}&+\frac{1}{3}&-\frac{1}{3}&+\frac{1}{3}
\rule{0mm}{6mm}\\
Y_3 &-\frac{1}{3}&+\frac{1}{3}&+\frac{1}{3}&-\frac{1}{3}
\rule{0mm}{6mm}\\
\end{array}
\end{math}
\end{center}
\end{table}

In \cite{APPB2007a} a conjecture was put forward that the electric charge 
$Q$ is just
an appropriately normalized operator $R^{tot}B_7$, evaluated
for the lowest level of $R$, i.e.:
\begin{equation}
\label{GMN}
Q=\frac{1}{6}(R_{lowest}+R^{\sigma})B_7=I_3+\frac{Y}{2}
\end{equation}
where, with $R_{lowest}=({\bf p}^2+{\bf x}^2)_{lowest}=3$, the eigenvalues of $I_3=B_7/2$ are $\pm 1/2$.
The above equation, {\it derived} here in a phase-space-related approach, 
 is known as the Gell-Mann-Nishijima relation (with $I_3$ known as weak isospin), and constitutes
a law of nature. 
With the help of Table \ref{table2}, 
it yields the charges of all eight leptons and quarks from a single
SM generation.

\subsection{Harari-Shupe rishons}
With the growing number of fundamental fermions,
the problem of understanding why they group into generations composed of eight
particles was addressed by many physicists.
The most widely cited proposal (over
320 citations) is due to Haim Harari and Michael Shupe \cite{Harari}.
The 
Harari-Shupe model describes the structure of a single SM
generation with the help of a composite model.
It builds all eight fermions of a single generation from 
only two spin-$1/2$ ``preons" $V$ and $T$
(or ``rishons" as Harari dubbed them),
 of charges
0 and +1/3 respectively. This is shown in Table \ref{table1}, where total 
charges and hypercharges of particles are also listed.
Note that rishons obey strange statistics, i.e. it is 
the states with ordered rishons (e.g. $VTT$, $TVT$, and $TTV$ ) which are to
correspond to the three colours of $SU(3)$ and which, consequently,
are deemed different.

The rishon model, though algebraically very economical, 
has several drawbacks, however.
These include, among others: the issue of preon confinement at extremely
small distance scales (when confronted
with the uncertainty principle), the apparent
absence of spin-3/2 fundamental particles, and the lack of explanation 
as to why the
ordering of three rishons is important and leads to $SU(3)$.

\begin{table}[t]
\caption{Rishon structure of leptons and quarks with weak
isospin $I_3=+1/2$}
\label{table1}
\begin{center}
\begin{math}
\begin{array}{c|cccc|cccc}
&~~~\nu_e~~~&~~u_R~~&~~u_G~~&~~u_B~~&~~e^+~~&
~~\bar{d}_R~~&~~\bar{d}_G~~&~~\bar{d}_B~~\\
\hline 
&VVV&VTT&TVT&TTV&TTT&TVV&VTV&VVT\rule{0mm}{5mm}\\
Q &0&+\frac{2}{3}&+\frac{2}{3}&+\frac{2}{3}&+1
&+\frac{1}{3}&+\frac{1}{3}&+\frac{1}{3}
\rule{0mm}{5mm}\\
Y &-1&+\frac{1}{3}&+\frac{1}{3}&+\frac{1}{3}&+1
&-\frac{1}{3}&-\frac{1}{3}&-\frac{1}{3}
\rule{0mm}{5mm}
\end{array}
\end{math}
\end{center}
\end{table}

\subsection{Preonless resolution of problems}
A comparison of Tables \ref{table2}, \ref{table1} (using Eq. (\ref{GMN}))
shows that the phase-space approach reproduces the main structure of the
Harari-Shupe model exactly.
In fact, it does not only that: it also solves the  three
problems of the rishon model.

Namely, the phase-space approach explains the structure of charge eigenvalues
without assuming {\it any} subparticle components of quarks and leptons. 
Therefore, there is no problem of ``where are
spin-3/2 fundamental fermions'', and there is no problem with preon confinement.
Furthermore, the strange statistics of rishons and its connection with SU(3) are
naturally explained.

One can readily understand the meaning of the ``ordered rishon structure'' (such
as $VTT$)
in phase-space terms. Thus, the position of rishon corresponds to one of three
directions in our macroscopic 3D space: $VTT$ corresponds therefore to the
partial hypercharge eigenvalue of $-1/3$ in direction $(x,p_x)$ and
to the same eigenvalue of $+1/3$ in both remaining
directions, $(y,p_y)$ and $(z,p_z)$.
Now, any discussion of rotations requires three directions, not just one.
Hence the concept of spin simply cannot be applied to a single rishon.
This is in line with Heisenberg's opinion \cite{Heisenberg} concerning the idea
of dividing matter again and again:
 ``...the antinomy of the smallest dimensions is  solved in particle physics
 in a very subtle manner, of which neither Kant nor the ancient
 philosophers could have thought: The word `dividing' loses its meaning''.

\section{Transformations in phase space}
\subsection{Genuine $SO(6)$ transformations}
In order to see the relation between quarks and leptons consider
\cite{APPB2007b,PLB2008} 
a transformation generated by $F^{\sigma}_{-2}$,
one of six ``genuine'' $SU(4)/SO(6)$ generators $F^{\sigma}_{\pm n}$ ($n=1,2,3$):
\begin{eqnarray}
F^{\sigma}_{-n}&=&
\frac{1}{2}\,(\sigma_0\otimes\sigma_n-\sigma_n\otimes\sigma_0)\,\sigma _3\\
F^{\sigma}_{+n}&=
&\frac{1}{2}\epsilon_{nkl}\sigma_k\otimes\sigma_l\otimes\sigma_3.
\end{eqnarray}
.
 
Under $F^{\sigma}_{-2}$-generated transformations one obtains:
\begin{eqnarray}
A'_k=A_1\cos
\phi-A_3\sin\phi&\phantom{xxxx}&B'_1=B_1\cos\phi+B_3\sin\phi\nonumber\\
A'_2=A_2~~~~~~~~~~~~~~~~~~~~~&&B'_2=B_2\\
A'_3=A_3\cos\phi+A_1\sin\phi\,&&B'_3=B_3\cos\phi-B_1\sin\phi\nonumber
\end{eqnarray}
i.e. ${\bf A}$ and ${\bf B}$ rotate in opposite senses.
For $\phi =\pm \pi/2$ one then gets:
\begin{equation}
Y=Y_1+Y_2+Y_3 \to Y'=-Y_3+Y_2-Y_1
\end{equation}
and consultation of Table \ref{table2} 
shows that lepton and quark \# 2 are exchanged, while the remaining two
quarks are left untouched. The same result is obtained when the analogous
transformation generated by $F^{\sigma}_{+2}$ is considered.

\subsection{Rotations in phase space}
When the corresponding transformations in phase space are considered, one
gets (for $F^{\sigma}_{-2}$-generated rotations):
\begin{eqnarray}
[x'_k,x'_l]=[p'_k,p'_l]&=&0\\
{}[x'_k,p'_l]&=&i\Delta_{kl}
\end{eqnarray}
with
\begin{equation}
\Delta=\left[
\begin{array}{ccc}
\cos 2\phi&0&\sin 2\phi\\
0&1&0\\
-\sin 2 \phi &0&\cos 2 \phi
\end{array}
\right].
\end{equation}
For the case of lepton-quark \# 2 interchange ($\phi = \pm \pi/2$) one then obtains
 (for both $F^{\sigma}_{-2}$- and $F^{\sigma}_{+2}$- generated rotations):
\begin{equation}
{\rm (quark)}~~~~~
\Delta=\left[
\begin{array}{ccc}
-1 &0&0\\
0&+1&0\\
0 &0&-1
\end{array}
\right]~~~~~
\leftrightarrow~~~~~
\Delta=\left[
\begin{array}{ccc}
1 &0&0\\
0&1&0\\
0 &0&1
\end{array}
\right]
~~~~~{\rm (lepton)}
\end{equation}
It may be said  therefore that quark is a lepton rotated in phase space.

Taking into account the remaining two types of genuine $SO(6)$
transformations we get the following four sets of generalized commutation
relations:
\begin{eqnarray}
{\rm lepton}~~~~~~~&&~~~{\rm quark~1}~~~~~~~{\rm quark~2}~~~~~~~{\rm quark~3}
\nonumber\\
{}[x_1,p_1]=i~~~~&&[x_1,p_1]=i~~~~[p_1,x_1]=i~~~~[p_1,x_1]=i\nonumber\\
\label{foursets}
{}[x_2,p_2]=i~~~~&&[p_2,x_2]=i~~~~[x_2,p_2]=i~~~~[p_2,x_2]=i\\
{}[x_3,p_3]=i~~~~&&[p_3,x_3]=i~~~~[p_3,x_3]=i~~~~[x_3,p_3]=i.\nonumber
\end{eqnarray}

\subsection{Reflections in phase space}
When a reflection in phase space (e.g.. $p'_k=p_k$, $x'_k=-x_k$, $i\to i$)
 is performed, the number of sets is doubled from four to eight, with $i$ in 
 Eqs (\ref{foursets}) changed to $-i$ (this corresponds to doublets of
 weak isospin and  is different from charge conjugation).
 Thus, we obtain 8 disjoint sectors, corresponding to 8 particles of a single
 generation of the Standard Model.

The fact that the three additional sets of commutation relations 
in Eq. (\ref{foursets}) are not 
rotationally invariant is in my opinion an asset of the approach.
Namely, the only condition that quarks must really fulfill is 
that it is the systems composed thereof
(i.e. mesons, baryons) that 
must be covariant under rotations. 
Thus, my conjecture is that quarks must conspire (see example in the next
section).

\section{Clifford algebra and mass}
The $U(1)\otimes SU(3)$ structure of (a half of) the Clifford algebra of nonrelativistic phase space
is shown in Table \ref{table3}. Here the even elements 
(linear combinations of $SU(4)/SO(6)$ generators and the unit element) 
with $I_3=+1/2$ (hence superscript `+')
 are given on the left,
while the odd elements (linear combinations of products of an odd number of $A_m$ and $B_n$) with left and right eigenvalues of
$I_{3l}=+1/2$ and $I_{3r}=-1/2$ are shown on the right. In the columns marked
$Y_l$ and $Y_r$, the left and right eigenvalues of $Y$ are given. 
A more detailed explanation of entries in this table may be found in \cite{JPA2009}.
\begin{table}[t]
\caption{Classification of Clifford algebra elements according to their 
$U(1)\otimes SU(3)$ properties}
\label{table3}
\begin{center}
\begin{math}
\begin{array}{ccccc|ccccc}
~~U(1)~&~SU(3)~&~{\rm elem}~&~Y_l~&~Y_r~&~~U(1)~&
~SU(3)~&~{\rm elem}~&~Y_l~&~Y_r~\\
\hline 
-2&3^*&H^+_{m0}&-1&+\frac{1}{3}&+1&3^*&U^{\dagger}_k&+\frac{1}{3}&+\frac{1}{3}\rule{0mm}{5mm}\\
+2&3&H^+_{0m}&+\frac{1}{3}&-1&-1&3&V_k&+\frac{1}{3}&-1\rule{0mm}{5mm}\\
0&8&F^+_a&+\frac{1}{3}&+\frac{1}{3}&-1&3&W_k&-1&-\frac{1}{3}\rule{0mm}{5mm}\\
0&1&Y^+_{-1}&-1&-1&+1&6&G_{\{kl\}}&+\frac{1}{3}&+\frac{1}{3}\rule{0mm}{5mm}\\
0&1&Y^+_{+\frac{1}{3}}&+\frac{1}{3}&+\frac{1}{3}&-3&1&G_0&-1&-1\rule{0mm}{5mm}
\end{array}
\end{math}
\end{center}
\end{table}

The algebraic counterpart of lepton mass should be odd 
(just like the odd $A_m$ is
associated with $p_m$) 
and is identified with the only odd
scalar element in Table \ref{table3}, i.e. with $G_0$.
The $F_{\pm 2}$-generated transformation from the lepton to the quark
sector changes $G_0$ into $G_{\{22\}}$, which is a member of the $SU(3)$ sextet,
and is not rotationally invariant.
However, the sum of the three quark mass terms, i.e. $G_{\{kk\}}$, is rotationally
invariant, just as the idea of quark conspiracy suggests.

\section{Summary}
The phase-space approach provides a possible theoretical explanation of
the
structure of a single generation of the Standard Model. The symmetry
 obtained involves $U(1)\otimes SU(3)$ and
resembles
the full  $U(1)\otimes SU(3) \otimes SU(2)_L$ symmetry group
of the Standard Model quite closely 
(in fact the $SU(2)$ partners
of $I_3=B_7/2$, 
i.e. $I_{1,2}=\sigma_0\otimes \sigma_0\otimes \sigma_{1,2}$, 
automatically do not commute with the 3D reflections, thus suggesting parity
violation). 
We have derived the Gell-Mann-Nishijima relation and reproduced
the structure of the Harari-Shupe model (visualized in Fig. 1), 
while evading its main problems related to the introduction of ``confined
preons''.
The existence of eight particles in a single SM generation
has been related to the $2^3=8$ possible sets of $[x_k,p_k]=\pm i$
commutation relations, with the $\pm$ sign adopted independently for each
direction in our 3D space.

The proposed aproach obviously raises many questions.
The phase-space-related modification of  the way in which the imaginary unit 
enters into our theories 
brings in the question of whether 
it is possible to extend the concept of the arena further, 
so that parity violation in weak interactions 
(together with the emergence of three quark-lepton generations and 
 the related appearance of the
$CP$-violating $i$) could be described in a more realistic way.
Then, there are other important questions such as the issue
of the emergence of
points, 
the construction of composite
systems, etc.
I hope to be able to address some of them in the future.\\

This work has been partially supported by the Polish Ministry of Science and
Higher Education research project No N N202 248135.
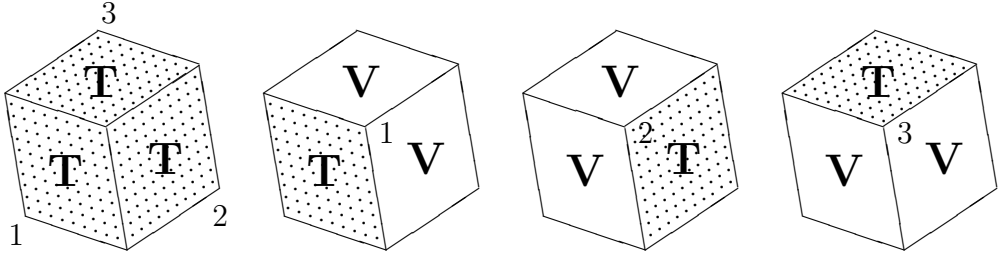
\begin{figure}
\begin{center}
\setlength{\unitlength}{0.7pt}
\begin{picture}(560,140)
\put(0,0){\begin{picture}(130,140)
\put(14,25.5){\line(-1,6){11}}
\put(69,7){\line(-1,6){11}}
\put(119,41){\line(-1,6){11}}
\put(3,92){\line(3,-1){55}}
\put(14,25.5){\line(3,-1){55}}
\put(69,7){\line(3,2){50}}
\put(58,74){\line(3,2){50}}
\put(3,92){\line(3,2){50}}
\put(53,126){\line(3,-1){55}}
\multiput(16,29.5)(-1.05,6.3){10}{\multiput(0,0)(6,-2){9}{\circle*{1}}}
\multiput(10,93)(4.8,3.2){10}{\multiput(0,0)(6,-2){9}{\circle*{1}}}
\multiput(72,15.3)(-1.05,6.3){10}{\multiput(0,0)(4.8,3.2){10}{\circle*{1}}}
\put(45,90){\makebox{\Large {\bf T}}}
\put(26,42){\makebox{\Large {\bf T}}}
\put(80,48){\makebox{\Large {\bf T}}}
\put(5,10){\makebox{1}}
\put(115,20){\makebox{2}}
\put(55,130){\makebox{3}}
\end{picture}}

\put(140,0){\begin{picture}(130,140)
\put(14,25.5){\line(-1,6){11}}
\put(69,7){\line(-1,6){11}}
\put(119,41){\line(-1,6){11}}
\put(3,92){\line(3,-1){55}}
\put(14,25.5){\line(3,-1){55}}
\put(69,7){\line(3,2){50}}
\put(58,74){\line(3,2){50}}
\put(3,92){\line(3,2){50}}
\put(53,126){\line(3,-1){55}}
\multiput(16,29.5)(-1.05,6.3){10}{\multiput(0,0)(6,-2){9}{\circle*{1}}}
\put(45,90){\makebox{\Large {\bf V}}}
\put(26,42){\makebox{\Large {\bf T}}}
\put(80,48){\makebox{\Large {\bf V}}}
\put(65,65){\makebox{1}}
\end{picture}}

\put(280,0){\begin{picture}(130,140)
\put(14,25.5){\line(-1,6){11}}
\put(69,7){\line(-1,6){11}}
\put(119,41){\line(-1,6){11}}
\put(3,92){\line(3,-1){55}}
\put(14,25.5){\line(3,-1){55}}
\put(69,7){\line(3,2){50}}
\put(58,74){\line(3,2){50}}
\put(3,92){\line(3,2){50}}
\put(53,126){\line(3,-1){55}}
\multiput(72,15.3)(-1.05,6.3){10}{\multiput(0,0)(4.8,3.2){10}{\circle*{1}}}
\put(45,90){\makebox{\Large {\bf V}}}
\put(26,42){\makebox{\Large {\bf V}}}
\put(80,48){\makebox{\Large {\bf T}}}
\put(65,65){\makebox{2}}
\end{picture}}

\put(420,0){\begin{picture}(130,140)
\put(14,25.5){\line(-1,6){11}}
\put(69,7){\line(-1,6){11}}
\put(119,41){\line(-1,6){11}}
\put(3,92){\line(3,-1){55}}
\put(14,25.5){\line(3,-1){55}}
\put(69,7){\line(3,2){50}}
\put(58,74){\line(3,2){50}}
\put(3,92){\line(3,2){50}}
\put(53,126){\line(3,-1){55}}
\multiput(10,93)(4.8,3.2){10}{\multiput(0,0)(6,-2){9}{\circle*{1}}}
\put(45,90){\makebox{\Large {\bf T}}}
\put(26,42){\makebox{\Large {\bf V}}}
\put(80,48){\makebox{\Large {\bf V}}}
\put(65,65){\makebox{3}}
\end{picture}}

\end{picture}
\end{center}
\caption{Four of the eight ``DICE1979'' corresponding 
to the Harari-Shupe model. All dice are identical when
their rotations are admitted.
  Each corner
corresponds to one particle of a single SM generation. Each face corresponds to
a rishon.
The three dice to the right (coloured quarks) show the leftmost die (lepton) rotated around axes
$1,2,3$ by $\pi$,
so that different corners (as marked) present themselves to the reader.}
\end{figure}

\vfill

\vfill

\end{document}